\documentclass[twocolumn,showpacs,preprintnumbers,amsmath,amssymb]{revtex4}
\usepackage{mathrsfs}
\usepackage{graphicx}% Include figure files
\usepackage{dcolumn}% Align table columns on decimal point
\usepackage{bm}% bold math

\def\deriv#1{\frac{\partial}{\partial#1}}
\newcommand{\half}{\frac{1}{2}}

\newcommand{\eg}{{\it e.g.}}

\begin{document}
\title{Tunneling spin current and spin diode behavior in bilayer system}

\author{Pei-Qing Jin  and You-Quan Li }
\address{Zhejiang Institute of Modern Physics and Department
of Physics, Zhejiang University, Hangzhou 310027, P. R. China}

\begin{abstract}
The coherent tunneling spin current in the bilayer system with
spin-orbit coupling is investigated. Based on the continuity-like
equations, we discuss the definition of the tunneling current and
show that the overlaps between wavefunctions for different layers
contribute to the tunneling current. We study the linear response
of the tunneling spin current to an in-plane electric field in the
presence of nonmagnetic impurities. The tunneling spin
conductivity we obtained presents a feature asymmetrical with
respect to the gate voltage when the strengthes of impurity
potentials are different in each layer.
\end{abstract}

%72.25.-b Spin polarized transport
%72.10.-d Theory of electronic transport; scattering mechanisms
%03.65.-w Quantum mechanics
%%85.75.-d Magnetoelectronics; spintronics: devices exploiting spin polarized transport or integrated magnetic fields
%%71.70.Ej Spin-orbit coupling, Zeeman and Stark splitting, Jahn-Teller effect
%%72.25.Dc Spin polarized transport in semiconductors

\received{\today}
\maketitle

\section{Introduction}
\label{intro}

The study of the tunneling process has a long history ever since the
foundation of quantum mechanics
for there are many applicable effects.
As an example, the coherent tunneling of cooper pairs
between weakly connected superconductors, known as the Josephson
effect, has been employed in the design of superconductor circuits
that are prospective
in quantum computation and quantum information~\cite{QI1,QI2}.
For the future information processing and storage technologies,
the emerging field spintronics~\cite{Daykonov, Wolf, Zutic} aims mainly
at coherent manipulation of spin degree of freedom and
controllable spin transport in solid states.
Owning to these anticipations, some attention has been
absorbed in the study of the coherent
spin tunneling process
either in the conventional Josephson
junctions~\cite{JC1,JC2,JC3} or
in the ferromagnetic tunnel junctions~\cite{Lee}.
However, as we are aware, the junction constituted by
a bilayer two-dimensional electron gases
has not been considered yet
although there exist versatile features in such systems.
For example, a resonantly enhanced tunneling was
reported in the bilayer quantum Hall system
when the layers are sufficiently close to each other~\cite{West}.
Moreover, the spin Hall conductivity,
which vanishes for arbitrarily small concentration of
nonmagnetic impurities in monolayer, was shown to have a magnification effect in
the bilayer electron system~\cite{Li-bilayer}.
We therefore investigate the coherent tunneling spin current
(TSC) in the bilayer system with spin-orbit coupling in this paper.
In our study, the tunneling is included in the unperturbed Hamiltonian
unlike the conventional tunneling Hamiltonian approach~\cite{Cohen}
where it is treated as a perturbation.

The present paper is organized as follows.
In Sec.~\ref{sec:definition},
we revisit the definition of the tunneling charge current and then
give the definition of the TSC
with the help of the continuity-like equations in bilayer systems.
In Sec.~\ref{sec:equal impurities}, we consider a twin layer system
in which the strengthes of impurity potentials in each layer are identical.
We study the linear response of the TSC to the in-plane electric field
and obtain the tunneling spin conductivity which exhibits sharp
cusps.
In Sec.~\ref{sec:diff impurities}, the influence on
the tunneling spin conductivity caused by the variation of the strengthes of
impurity potentials in different layer is considered.
The expression of the Green's function is extended so as to
incorporate such influences.
We find that the difference between the strengthes of impurity potentials
in the two layers
gives rise to the asymmetrical feature of
the tunneling
spin current with respect to the gate voltage.
Finally, a brief summary is given in
Sec.~\ref{sec:summary} and some concrete expressions are written out
in the Appendix.

\section{Definition of coherent tunneling spin
current}\label{sec:definition}

In order to properly define the coherent TSC,
let us first recall the definition of the tunneling charge
current across a junction.
Such a current is generated by electrons tunneling
from one side of the junction to the other side
due to the imbalance of the chemical potentials produced by
an applied gate voltage $V$.
The bilayer system considered in this paper resembles
the tunnel junction and its Hamiltonian is given by
\begin{eqnarray}
H = \frac{p^2}{2m} I + eV \tau_z + \beta\tau_x,
\end{eqnarray}
where $\beta$ is the tunneling strength, $I$ and $\tau_a$
($a=x,y,z$) denote the unit matrix and Pauli matrices in the layer
representation, respectively. The wavefunction of the system is
expressed as $\Psi = (\psi^{}_f, \psi^{}_b)^T$. From the
Schr\"{o}dinger equation, we can obtain the continuity-like equations
for the densities
$\rho^{}_{\ell}=\psi^{\dag}_{\ell}\psi^{}_{\ell}$, namely,
\begin{eqnarray}\label{eq:continuity-charge}
\frac{\partial \rho^{}_{\!f}}{\partial t}
 + \frac{\partial j^{}_{\!fi}}{\partial x_i}
 &=& \frac{i\beta}{\hbar}
 (\psi^{\dag}_{b}\psi^{}_{\!f}-h.c.),
                \nonumber \\
\frac{\partial \rho^{}_{b}}{\partial t}
 + \frac{\partial j^{}_{bi}}{\partial x_i}
 &=& \frac{i\beta}{\hbar}
 (\psi^{\dag}_{\!f}\psi^{}_{b}-h.c.),
\end{eqnarray}
with $j^{}_{\ell i}= -\frac{i\hbar}{2m}
(\psi^{\dag}_{\ell}\frac{\partial}{\partial x_i}
\psi^{}_{\ell}-h.c.)$ and $h.c.$ refers to the hermitian conjugation
since $\psi^{}_{\ell}$ should be in a two-component form
$\psi^{}_{\ell}=(\psi^{}_{\ell\uparrow},\psi^{}_{\ell\downarrow})^T$
if the spin degree of freedom is taken into account.
Throughout this paper, the subscripts
$i$ represent the components of a quantity in the spatial space and
$\ell$ label the quantities of the front layer with $\ell=f$ or the
back layer with $\ell=b$.
The nonvanishing terms on the right hand sides of
Eqs.~(\ref{eq:continuity-charge}), which are caused by the tunneling
between layers, indicate the overlap between the wavefunctions for
different layers.

Integrating the first equation of Eqs.~(\ref{eq:continuity-charge})
over the domain of the front
layer ${\mathcal D}_f$, we have
\begin{eqnarray}\label{eq:def-charge}
I^{}_T = -\frac{d N^{}_{\!f}}{dt} =\! \int\! j^{}_{fz} ~d A -
\frac{i\beta}{\hbar}\int_{{\mathcal D}_f}
 (\psi^{\dag}_{b}\psi^{}_{\!f}-h.c.) ~dV,
\end{eqnarray}
where $A$ denotes the area of the layer and $N^{}_{f} =
\int_{{\mathcal D}^{}_{\!f}}\rho^{}_{\! f}~dV$ the electron number in
the front layer. Here we focus on the tunneling-relevant
direction, saying the $z$-direction, and assume that there is no
in-plane current flowing out of each layer.
Equation~(\ref{eq:def-charge}) demonstrates that the rate of the
change of the electron number in the front layer, defined as
the tunneling current, contains not only  the conventional
contribution $\int\! j^{}_{fz} ~d A$, but also the overlap between
the wavefunctions for different layers. The orthogonality and
completeness of the states on each side of the junction were
discussed~\cite{state1,state2,state3}. Then
Eq.~(\ref{eq:def-charge}) manifests, from another point of view,
why the tunneling current is conventionally evaluated by the rate
of the change of the electron number on one side of a junction.

Now we are in the position to consider the TSC in
the bilayer system with Rashba-type spin-orbit coupling.
A natural definition of the TSC density in a system
should be the rate of change of the spin density
$\vec S_{\ell} = \psi^\dag_{\ell}~\vec s~\psi^{}_{\ell}$ in the
$\ell$-layer where $\vec s = \frac{\hbar}{2}\vec \sigma$ are the
spin operators.
Here and hereafter the overhead arrow represents that
the quantity is a vector in the spin space.
The continuity-like equations
for the spin density $\vec S_{\ell}$ in the system with SU(2) gauge
potentials $\vec{\mathcal A}_{0}$ and $\vec{\mathcal
A}_{i}$~\cite{Li-current} have been obtained in our previous
paper~\cite{Li-bilayer}, which read
\begin{eqnarray}\label{eq:continuity-spin}
\bigl(\deriv{t}
   -\eta\vec{\mathcal A}_{f 0}\times\bigr)\vec S_f
   +\bigl(\deriv{x_i}
     +\eta\vec{\mathcal A}_{f i}
       \times\bigr)\vec{J}^{}_{f i}
              \hspace{10mm} \nonumber \\
=\frac{i \beta}{\hbar}
  (\psi^\dag_{ b} ~\vec s~ \psi^{}_{\!f} -h.c.),
    \nonumber\\
  \bigl(\deriv{t}
  -\eta\vec{\mathcal A}_{b 0}\times\bigr)\vec S_b
   +\bigl(\deriv{x_i}
    +\eta\vec{\mathcal A}_{b i}
      \times\bigr)\vec{J}^{}_{b i}
              \hspace{11mm} \nonumber \\
=\frac{i \beta}{\hbar} (\psi^\dag_{\!f}~\vec s~ \psi^{}_{b}
    -h.c.),
\end{eqnarray}
with $\vec J^{}_{\ell i} = \textrm{Re} (\psi^\dag_{\ell}~ \half
\{v_i,\vec s\} ~\!\psi^{}_{\ell})$ where $v_i$ is the velocity
operator and the curl brackets denote the anti-commutation relation.
Let $\alpha_1$ and $\alpha_2$ stand for
the spin-orbit coupling strengthes in the front and back layers,
respectively, and then we have
$\vec{\mathcal A}_{f x}
\!=\!\frac{2m}{\eta^2}(0,~\alpha_1~,0)$, $\vec{\mathcal A}_{f y}
\!=\!-\frac{2m}{\eta^2}(\alpha_1,~0~ ,0)$, $\vec{\mathcal A}_{b x}
\!=\!\frac{2m}{\eta^2}(0,~\alpha_2~,0)$, $\vec{\mathcal A}_{b y}
\!=\!-\frac{2m}{\eta^2}(\alpha_2,~0~ ,0)$ and $\vec{\mathcal A}_{f
z}\!=\!\vec{\mathcal A}_{f 0}\!=\! \vec{\mathcal A}_{b
z}\!=\!\vec{\mathcal A}_{b 0}\!=\!0$ with $\eta=\hbar$.

Since the tunneling being considered is spin-independent,
only a gate voltage can not induce a nonvanishing TSC.
As we known, an in-plane electric field is applied to
drive the spin Hall current and the basic relation between
the spin current and the electric field is given by~\cite{SHE}
\begin{eqnarray}\label{eq:J-E relation}
 J^a_i = \epsilon^{}_{aij} \sigma^{}_s E^{}_j,
\end{eqnarray}
where $\sigma^{}_s$ is the spin Hall conductivity and
$\epsilon^{}_{aij}$ the totally antisymmetric tensor. It
demonstrates that the flow direction and the spin-polarization
direction of the current as well as the direction of the electric
field are always perpendicular to each other.
Since the tunneling is with respect to the $z$-direction
and the in-plane electric field is along the $x$-direction,
we focus on the component of the TSC polarized in the $y$-direction.
From Eq.~(\ref{eq:continuity-spin}), we have
\begin{eqnarray}\label{eq:def-spin}
-\frac{\partial S^y_{\!f}}{\partial t}
 = (\frac{\partial}{\partial z}
 +\eta{\mathcal A}^x_{\!f y})J^y_{\!f z}
 -\frac{i \beta}{\hbar}
  (\psi^\dag_{ b} s^y \psi^{}_{\!f} -h.c.),
\end{eqnarray}
where we have taken that the nonvanishing components of
the spin current are $J^y_{\!fz}=-J^z_{\!fy}$ as the electric
field is along the $x$-direction. The last term on
the right hand side of
Eq.~(\ref{eq:def-spin}) can be regarded as the overlap between the
eigenfunctions of $s^y$. Unlike the case of the tunneling charge
current, the contributions of the states with spin parallel
and anti-parallel to the $y$-axis have opposite signs. The
covariant derivative $\frac{\partial}{\partial z}+\eta{\mathcal
A}^x_{\!f y}$ is in place of the conventional derivative, which
indicates that the spin precession due to the spin-orbit coupling
leads to additional contribution to the TSC.

\section{Tunneling Spin Current in twin-layer system}\label{sec:equal impurities}

In the previous section, the definition
of the TSC in bilayer systems with
Rashba-type spin-orbit coupling in each layer
has been discussed by taking account of electrons' coherent tunneling.
Unlike the tunneling Hamiltonian approach~\cite{Cohen,Mahan} in the study
of the tunneling charge current, here we take
the tunneling term in the unperturbed Hamiltonian.
Since the tunneling strength is not necessarily weak
in our approach, we are able to attain more
information beyond the perturbation approach.
In the present approach, the unperturbed Hamiltonian in the second
quantization form is given by
\begin{eqnarray}\label{Hamiltonian}
& H = \sum_{\mathbf k}
C^\dag_{\mathbf k}
 \Bigl\{
\left(%
\begin{array}{cc}
  \!\frac{\hbar^2k^2}{2m} + e V\!\! & \!\!\beta\!\! \\
  \!\!\beta\!\!   & \frac{\hbar^2k^2}{2m}- e V \\
\end{array}%
\right)\otimes I \hspace{6mm}
         \nonumber  \\
  &\hspace{3mm} +\Bigl(\alpha_+I + \alpha_-\tau_z
     \Bigr)\otimes(k_y\sigma^x \!-\! k_x\sigma^y)
     \Bigr\} C_{\mathbf k} + \hat V_\text{im}.
\end{eqnarray}
We have adopted notations $\alpha_+
=(\alpha_1+\alpha_2)/2$, $\alpha_- =(\alpha_1-\alpha_2)/2$ and
$C^\dag_\mathbf k \equiv (c^\dag_{\mathbf
k,f\uparrow},c^\dag_{\mathbf k,f\downarrow}, c^\dag_{\mathbf
k,b\uparrow},c^\dag_{\mathbf k,b\downarrow})$
with $c^\dag_{\mathbf k,\ell\uparrow(\downarrow)}$
being the creation operator for a
spin-up (spin-down) electron in the $\ell$-layer.
Throughout the paper,
the boldface of a quantity manifests that it is a two-dimensional
vector, \eg, $\mathbf k = (k_x,k_y)$.
The last term in the Hamiltonian characterizes the interaction
between electrons and impurities.
The influence
of nonmagnetic~\cite{disnon1,disnon2,disnon3,Dimitrova}
and magnetic~\cite{dismag1,dismag2} impurities
on the spin Hall conductivity
were discussed in monolayer systems.

For simplicity,
we consider the nonmagnetic impurities in both layers
are all alike in this section.
Thus the potential energy of impurities is
given by $\hat V_\text{im} = u\sum_i \delta(\mathbf r-\mathbf R_i)$ with
$u$ being the strength and $\mathbf R_i$ the position of the
impurity. Here we assume the interaction strength $u$ is weak so that the Born
approximation~\cite{Haug} is applicable.

In the chiral representation, the Hamiltonian is diagonalized by a
unitary matrix $U$ with eigenenergies
\begin{eqnarray*}
\varepsilon_1 = \epsilon_k- \alpha_+ k
               -\lambda_{12},
                         \hspace{5mm}
\varepsilon_2 = \epsilon_k - \alpha_+ k
               +\lambda_{12},
                         \nonumber \\
\varepsilon_3 = \epsilon_k + \alpha_+ k
               -\lambda_{34},
                         \hspace{5mm}
\varepsilon_4 = \epsilon_k + \alpha_+ k
               +\lambda_{34},
\end{eqnarray*}
with $\epsilon_k = \frac{\hbar^2 k^2}{2m}$, $\lambda_{12}=\sqrt{(e
V-\alpha_-k)^2+\beta^2}$ and $\lambda_{34}=\sqrt{(e
V+\alpha_-k)^2+\beta^2}$.
Hereafter, we set $\hbar=1$ for
simplicity.
The free retarded Green's function in this chiral representation
is given by $G^{R}_{0(ch)} (\mathbf k,\omega) =
\textrm{diag}((\omega-\varepsilon_1+i\eta)^{-1},
(\omega-\varepsilon_2+i\eta)^{-1},(\omega-\varepsilon_3+i\eta)^{-1},
(\omega-\varepsilon_4+i\eta)^{-1})$ with $i\eta$ being an
infinitesimal quantity. And the Green's function in the original
representation is related to that in the chiral representation by the unitary
transformation, namely, $G(\mathbf k,\omega)=U(\mathbf k)
~G_{(ch)}(\mathbf k,\omega)~U^\dag(\mathbf k)$.

As a macroscopic quantity, the TSC is expected not to be affected by
the details of impurity location $\{\mathbf R_i \}$.
We assume there is no correlation between impurities and employ the
impurity averaging techniques and the diagrammatic method
~\cite{Abrikosov,Haug}.
A physical quantity $Q$ for the whole system
is obtained by taking average over impurities'
configuration, namely, $\overline Q=\langle Q(\{\mathbf
R_i\})\rangle_\text{im}=\Pi_i\int\frac{d\mathbf R_i}{A} Q(\{\mathbf
R_i\})$. In the Born approximation, the averaged retarded Green's
function $\overline {G^R}(\mathbf k,\omega)$ satisfies the Dyson
equation
\begin{eqnarray}\label{eq:Dyson-equal}
\overline {G^R} (\mathbf k,\omega) &=&
 G^R_0 (\mathbf k,\omega) +
 G^R_0 (\mathbf k,\omega) \Bigl( un_\text{im}
                    \nonumber \\
&& +\frac{u^2 n_\text{im}}{A}\sum_{\mathbf q}
 \overline{G^R}(\mathbf q,\omega) \Bigr)
 \overline{G^R}(\mathbf k,\omega),
\end{eqnarray}
where $n_\text{im}$ stands for the impurity concentration.
The above equation has a self-consistent solution,
$\overline{G^{R}_{(ch)}}(\mathbf k,\omega)
=\textrm{diag}(g_1,g_2,g_3,g_4)$ with
$g_j=(\omega-\varepsilon_j-un_\text{im}+\frac{i}{2\tau})^{-1}$
where the subscript $j$ runs from $1$ to $4$.
Here $\tau=(2\pi u^2 n_\text{im}
N^{}_F)^{-1}$ is the momentum-relaxation time and $N^{}_F$ the density
of states at the Fermi surface. Note that the averaged retarded
Green's function is diagonal in the chiral representation.
We will see in the next section
that the difference between the strengthes of impurity potentials
in two layers results in the emergence of off-diagonal elements
in $\overline{G^R_{(ch)}}$.

The TSC defined in the previous section can be expressed in a
symmetric form with respect to both layers, namely, the difference
between the rate of change of the spin operator in each layer
\begin{eqnarray}\label{eq:TSC}
J^y_z \!=\! -\half\langle\frac{d \hat S^y_-}{dt} \rangle
 \!=\!- \half\langle \frac{d}{dt}\sum_{\mathbf k} C^\dag_{\mathbf
k} (\tau_z\otimes s^y) C_{\mathbf k} \rangle.
\end{eqnarray}
Its linear response to the external in-plane electric field, the
tunneling spin conductivity $\overline{\sigma^y}$, can be obtained
by using the Kubo formula
\begin{eqnarray}
\overline{\sigma^y}(\omega) = \frac{1}{2\pi\omega A}
 \int d\omega_1     \textrm{Tr} \{ n^{}_F(\omega+\omega_1)
           \hspace{22mm}     \nonumber \\
 \times\overline{[G^R(\omega+\omega_1)-G^A(\omega+\omega_1)] \hat j_e
 G^A(\omega_1)\hat j^y_z}
            \hspace{6mm}     \nonumber \\
             + n^{}_F(\omega_1)
  \overline{[G^R(\omega_1) \!-\! G^A(\omega_1)]
  \hat j^y_z G^R(\omega+\omega_1)\hat j_e}
  \},
\end{eqnarray}
where $\hat j^y_z
 \!=\! \half [(\alpha_+\tau_z+\alpha_- I)
 \otimes k_y\sigma_z-\beta\tau_y\otimes\sigma_y]$
and $\textrm{Tr}$ refers to the trace taken over the spin indices as
well as the summation over the momentum. $G^A$ is the advanced
Green's function, $n^{}_F$ the Fermi distribution function and
$\hat j_e=e\hat v_x$ the charge current operator.

In the uncrossing approximation~\cite{Dimitrova}, the dc tunneling
spin conductivity $\overline{\sigma^y}$ at zero temperature is
calculated as the sum of contributions of the one-loop diagram
$\overline{\sigma^y_0}$ and a series of ladder diagrams
$\overline{\sigma^y_L}$.
The former is given by
$\overline{\sigma^y_0}=-\frac{1}{2\pi A}
 \textrm{Tr} (\overline{G^R}(\mathbf k)\hat j_e(\mathbf k)
 \overline{G^A}(\mathbf k)\hat j^y_z(\mathbf k) )$,
and denoted diagrammatically as

%\begin{figure}[h]
\begin{center}
\begin{picture}(80,50)(-16,-25)
 \put(-59,-2){$\overline{\sigma^y_0}=-\displaystyle\frac{1}{2\pi A}$}
 \put(-2,-2){$\hat j^y_z$}
 \put(15,0){\circle*{8}}
 \qbezier(15,0)(15,17)(40,17)
 \put(37,20){$\overline{G^A}$}
 \qbezier(40,17)(61,17)(64,4)
 \put(65,0){\circle{8}}
 \put(73,-2){$\hat j_e ~.$}
 \qbezier(15,0)(15,-17)(40,-17)
 \put(36,-30){$\overline{G^R}$}
 \qbezier(40,-17)(61,-17)(64,-4)
\end{picture}
\end{center}
%\end{figure}
A direct calculation gives rise to
\begin{eqnarray}\label{eq:conduct-non}
\overline{\sigma^y_0} = -2\alpha_-\sum_{\mathbf k} \sin^2\varphi
           \hspace{49mm}      \nonumber \\
 ~\times\textrm{Im}\Bigl[\frac{\beta^2}{\lambda_{12}} a_1 g_2
 +\frac{\beta^2}{\lambda_{34}} a_3 g_4
 +\alpha_+k (a_1+a_2)(g_3+g_4)
           \hspace{3.5mm}      \nonumber \\
 +\frac{\beta^2+(\alpha^2_++\alpha^2_-)k^2
 -eV\frac{\alpha^2_++\alpha^2_-}{\alpha_-}k}{2\lambda_{12}}
 (a_1-a_2)(g_3+g_4)
                 \nonumber \\
 -\frac{\beta^2+(\alpha^2_++\alpha^2_-)k^2
 +eV\frac{\alpha^2_++\alpha^2_-}{\alpha_-}k}{2\lambda_{34}}
 (a_1+a_2)(g_3-g_4)
                 \nonumber \\
  + \frac{(e^2V^2 \!-\! \beta^2 \!-\! \alpha^2_-k^2)\alpha_+k}
 {\lambda_{12}\lambda_{34}} (a_1-a_2)(g_3-g_4) \Bigr],
           \hspace{12mm}
\end{eqnarray}
where the polar coordinates $\mathbf k=(k\cos\varphi,k\sin\varphi)$
is adopted and
$a_i$ are the matrix elements of the advanced Green's function
in the chiral representation.

\begin{figure}[h]
\includegraphics[width=70mm]{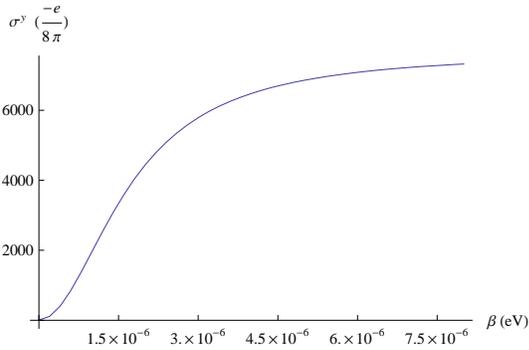}
\caption{(Color online) $\overline{\sigma^y}$ as a function of
$\beta$. The parameters are given by $V=0$, $\alpha_+ = 5.5\times
10^{-14} \textrm{eV~m}$, $\alpha_- = 0.45\times 10^{-14}
\textrm{eV~m}$, $\mu = 0.1~\textrm{eV}$, $\tau = 660~\textrm{fs}$
and the effective mass $m=0.065~m_0$ as in GaAs with $m_0$ being the
mass of the free electron. }\label{fig:tun}
\end{figure}

In the limit of large Fermi circle, $\mu \gg \Delta_{ij},1/\tau$
with $\Delta_{ij}=\varepsilon_i-\varepsilon_j$, the vertex
correction to the conductivity $\overline{\sigma^y_L}$ is dominated by
the terms with one advanced and one retarded Green's
function~\cite{Dimitrova}, which can be expressed diagrammatically as
\begin{center}
\begin{picture}(150,56)(38,-25)
 \put(12,-2){$\,\hat j^y_z$}
 \put(25,0){\circle*{6}}
 \qbezier(25,0)(25,17)(50,17)
 \put(47,20){$\overline{G^A}$}
 \qbezier(50,17)(71,17)(74,3)
 \put(75,0){\circle{6}}
 \put(82,-2){$\hat j_e$}
 \qbezier(25,0)(25,-17)(50,-17)
 \put(46,-29){$\overline{G^R}$}
 \qbezier(50,-17)(71,-17)(74,-3)
 \multiput(36,13)(0,-4){8}{\line(0,1){2}}
 \put(93,-2){$+$}
 \put(106,-2){$\hat j^y_z$}
 \put(120,0){\circle*{6}}
 \qbezier(120,0)(120,17)(145,17)
 \put(141,20){$\overline{G^A}$}
 \qbezier(145,17)(166,17)(169,3)
 \put(170,0){\circle{6}}
 \put(177,-2){$\hat j_e$}
 \qbezier(120,0)(120,-17)(145,-17)
 \put(140,-29){$\overline{G^R}$}
 \qbezier(145,-17)(166,-17)(169,-3)
 \multiput(130.5,13)(0,-4){8}{\line(0,1){2}}
 \multiput(143,15)(0,-4){9}{\line(0,1){2}}
 \put(187,-2){$+\cdots.$}
\end{picture}
\end{center}
If we introduce a vertex ${\mathcal J}^y$ which is the sum of vertex
corrections to $\hat j^y_z$,
\begin{center}
\begin{picture}(130,30)(15,-5)
 \put(0,-2){$\mathcal{J}^y~\equiv~$}
 \put(30,-2){$\hat j^y_z$}
 \put(45,0){\circle*{6}}
 \qbezier(45,0)(45,11)(57,13)
 \qbezier(45,0)(45,-11)(57,-13)
 \multiput(57,11)(0,-4){7}{\line(0,1){2}}
 \put(67,-2){$+$}
 \put(80,-2){$\hat j^y_z$}
 \put(95,0){\circle*{6}}
 \qbezier(95,0)(95,15)(118,15)
 \qbezier(95,0)(95,-15)(118,-15)
 \multiput(106,11)(0,-4){7}{\line(0,1){2}}
 \multiput(118,13)(0,-4){8}{\line(0,1){2}}
 \put(130,-2){$+~\cdots$,}
\end{picture}\vspace{3mm}
\end{center}
then $\overline{\sigma^y_L}$ can be expressed in a
compact form,
\begin{eqnarray}
\overline{\sigma^y_L}=\frac{-1}{2\pi A}
 \textrm{Tr} (\overline{G^R}(\mathbf k)\hat j_e(\mathbf k)
 \overline{G^A}(\mathbf k)\mathcal {J}^y ),
\end{eqnarray}
and the explicit expression of $\overline{\sigma^y_L}$ in terms of
Green's functions is given in the Appendix. The momentum independent
$\mathcal{J}^y$ can be obtained from the transfer matrix equation
\begin{eqnarray}\label{eq:transfer}
\mathcal J^y = \frac{u^2 n_\text{im}}{A}
 \sum_{\mathbf q} \overline{G^A}(\mathbf q)
 (\hat j^y_z(\mathbf q) + \mathcal J^y)
 \overline{G^R}(\mathbf q).
\end{eqnarray}
Here the summation over the momentum can be evaluated by taking it
as integration in the limit of large Fermi circle.
Then we obtain the tunneling spin conductivity $\overline{\sigma^y}$
which is plotted as a function of $\beta$ in Fig.~\ref{fig:tun}.
We can see that the TSC vanishes when the tunneling is absent.

\begin{figure}[h]
\includegraphics[width=88mm]{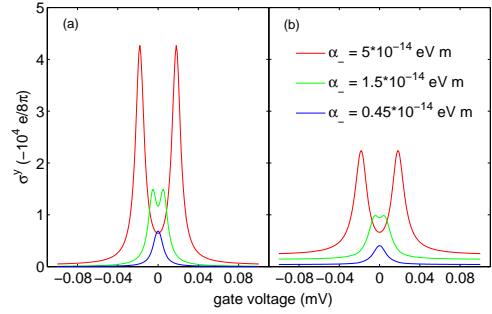}
\caption{(Color online) Tunneling spin conductivities
$\overline{\sigma^y}$ are plotted as functions of the gate voltage
$V$ with different values of $\alpha_-$ given in the legend. The
tunneling strength is $\beta=5\times10^{-6} \textrm{eV}$, the
momentum relaxation time are $\tau = 660 \textrm{fs}$ in panel (a)
and $\tau = 66 \textrm{fs}$ in panel (b) while the other parameters are
the same as those in Fig.~\ref{fig:tun}. }\label{fig:same3d}
\end{figure}

We plot $\overline{\sigma^y}$ as a function of the applied gate
voltage $V$ with different $\alpha_-$ and momentum relaxation time
in Fig.~\ref{fig:same3d}.
It is shown that $\overline{\sigma^y}$ is
an even function of the gate voltage.
For small $\alpha_-$, a peak in $\overline{\sigma^y}$
appears around $V=0$.
As $\alpha_-$ increases, this peak begins to split
and its height increases.
Additionally,
the nonmagnetic impurities tend to suppress these
peaks, as shown in Fig.~\ref{fig:same3d}(b) where the momentum
relaxation time is a tenth part of that in Fig.~\ref{fig:same3d}(a).

\begin{figure}[h]
\includegraphics[width=80mm]{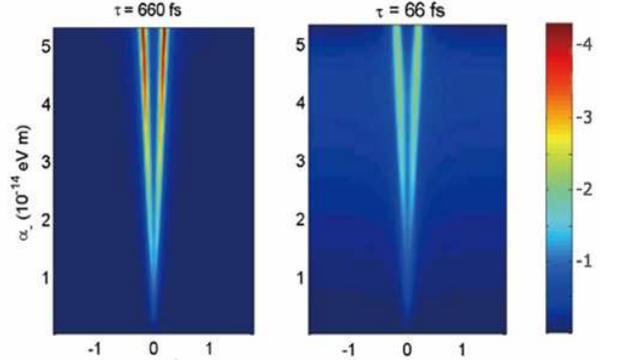}
\caption{(Color online) Contour plot of $\overline{\sigma^y}$ which
is a function of both the gate voltage and $\alpha_-$. The color bar
represents the value of $\overline{\sigma^y}$ in unit of $10^4
e/8\pi$. }\label{fig:same2d}
\end{figure}

Figure~\ref{fig:same2d} shows the contour-plot projections of
$\overline{\sigma^y}$ which are plotted as a function of both the
gate voltage and $\alpha_-$. These figures indicate that the
resonant peaks in $\overline{\sigma^y}$ begin to converge as
$\alpha_-$ decreases.

\section{Tunneling spin current asymmetrical
to gate voltage}\label{sec:diff impurities}

In realistic samples, the strengthes of impurity potentials in those two layers
may not happen to be identical.
We thus introduce $u^{}_{\!f}$ and
$u^{}_{b}$ to denote the strengthes of impurity potentials in the front and
back layers, respectively. Accordingly, the interaction between
electrons and impurities is given by
\begin{eqnarray}
\hat V_\text{im}= \sum_i
\delta(\mathbf r-\mathbf R_i) \left(
  \begin{array}{cc}
    u^{}_{\!f}  & 0 \\
    0 & u^{}_{b}  \\
  \end{array}
\right)\otimes I.
\end{eqnarray}
Note that the unit matrix in the layer space is no longer
appropriate in the present case.
This implies that the line which refers to the impurity potential
in the Feynman diagram (see Fig.~\ref{fig:matrix})
represents a matrix $\textrm{diag}(u^{}_{f}/A, u^{}_b/A)$
rather than a number $u/A$.
\begin{figure}[h]
\begin{center}
\begin{picture}(150,30)(15,-4)
 \put(0,-2){\vector(2,0){30}}
 \multiput(13.5,-2)(0,3){7}{\circle*{0.5}}
 \put(8,16){\line(5,2){11}}
 \put(19,16){\line(-5,2){11}}
 \put(12,-12){$\mathbf k$}
 \put(35,7){$=$}
 \put(50,7){$N~G_0(\mathbf k)$}
 \put(90,7){$\left(
               \begin{array}{cc}
                \displaystyle\frac{u^{}_{\!f}}{A} & 0 \\[2mm]
                 0 & \displaystyle\frac{u^{}_b}{A}
               \end{array}
             \right)$}
 \put(145,7){$G_0(\mathbf k)$}
\end{picture}
\end{center}
\caption{The first order Feynman diagram in the expansion of the
impurity-averaged Green's function where the momentum is conserved
at the vertex. The vector refers to the free Green's function. And
the dotted line represents the impurity potential which is in the
matrix form for the unequal strengthes of impurity potentials between layers.
}\label{fig:matrix}
\end{figure}
Accordingly, the Dyson equation for $\overline {G^R}$ is then
written as
\begin{eqnarray}\label{eq:Dyson-diff}
\overline {G^R} (\mathbf k,\omega) =
 G^R_0 (\mathbf k,\omega) +
 G^R_0 (\mathbf k,\omega) N
 \Big[
 \left(
  \begin{array}{cc}
    \frac{u^{}_{\!f}}{A} & 0 \\
    0 & \frac{u^{}_{b}}{A} \\
  \end{array}
 \right)
          \hspace{3mm} \nonumber \\[1mm]
 +\sum_{\mathbf q}
 \left(
  \begin{array}{cc}
    \frac{u^{}_{\!f}}{A} & 0 \\
    0 & \frac{u^{}_{b}}{A} \\
  \end{array}
\right)
 \overline{G^R}(\mathbf q,\omega)
 \left(
  \begin{array}{cc}
    \frac{u^{}_{\! f}}{A} & 0 \\
    0 & \frac{u^{}_{b}}{A} \\
  \end{array}
\right) \Bigr]
 \overline{G^R}(\mathbf k,\omega).
\end{eqnarray}
We obtain a self-consistent solution for the above equation
\begin{eqnarray} \overline{G^R_{(ch)}}=
 \left(
   \begin{array}{cccc}
     R_{11} & R_{\!f} & 0 & 0 \\
     R_{\!f} & R_{22} & 0 & 0 \\
     0 & 0 & R_{33} & R_{b} \\
     0 & 0 & R_{b} & R_{44} \\
   \end{array}
 \right),
\end{eqnarray}
with nonvanishing off-diagonal elements even in the chiral representation.
The explicit expressions for those matrix elements are given in the
Appendix.
The momentum relaxation time for the front and back layers
are simply given by $\tau^{}_{\!f}=(2\pi u^2_{\!f} n_\text{im} N^{}_F)^{-1}$
and $\tau^{}_{b}=(2\pi u^2_{b} n_\text{im} N^{}_F)^{-1}$, respectively.

The difference between the strengthes of impurity potentials in two layers
not only modifies the diagonal elements of $\overline{G^R_{(ch)}}$
but also requires the appearance of off-diagonal elements inevitably.
Now the contributions to $\overline{\sigma^y}$
by the diagonal elements can be directly obtained
by replacing $g_i$ and $a_i$, respectively, by $R_{ii}$ and $A_{ii}$
in Eq.~(\ref{eq:conduct-non}) and Eq.~(\ref{eq:conduct-cross}).
Here $A_{ii}=R_{ii}^*$
denote the diagonal elements of the averaged
advanced Green's function in the chiral representation.
And the contributions
to $\overline{\sigma^y_0}$ given by the off-diagonal elements reads
\begin{eqnarray}
\overline{\sigma^y_{0,d}} =-
 2\beta\sum_{\mathbf k} \sin^2\!\varphi~
 \textrm{Im}\Bigl[
                   \hspace{47mm}     \nonumber \\
 \times R_{\textrm{f}} ((A_{11}\!-\!A_{22})
 (\frac{k}{m}\!-\!\alpha_+)
 +(A_{11}\!+\!A_{22})
 \frac{\alpha_-(eV\!-\!\alpha_-k)}{\lambda_{12}}
                          \nonumber \\
 +(A_{33}\!+\!A_{44})
 \frac{eV \alpha_-\!+\!\alpha^2_+k}{\lambda_{12}}
 -(A_{33}\!-\!A_{44})
 \frac{\alpha_+\lambda_{34}}{\lambda_{12}})
                   \hspace{10mm}    \nonumber \\
 -R_{\textrm{b}} ((A_{33}\!-\!A_{44})
 (\frac{k}{m}\!+\!\alpha_+)
 -(A_{33}\!+\!A_{44})
 \frac{\alpha_-(eV\!+\!\alpha_-k)}{\lambda_{34}}
                   \hspace{-1mm}     \nonumber \\
 -(A_{11}\!+\!A_{22})
 \frac{eV \alpha_-\!-\!\alpha^2_+k}{\lambda_{34}}
 +(A_{11}\!-\!A_{22})
 \frac{\alpha_+\lambda_{12}}{\lambda_{34}})\Bigr].
                  \hspace{7.5mm}    \nonumber
\end{eqnarray}
The transfer matrix equation for the vertex $\mathcal J^y$ is also
modified as
\begin{eqnarray}
\mathcal J^y = N \sum_{\mathbf q}
 \left(\begin{array}{cc}
                 \frac{u_{\!f}}{A} & 0 \\
                 0 & \frac{u_b}{A} \\
               \end{array} \right)
 \overline{G^A}(\mathbf q)
            \hspace{14mm}     \nonumber \\
 \times(\hat j^y_z(\mathbf q) + \mathcal J^y)~
 \overline{G^R}(\mathbf q)
 \left(\begin{array}{cc}
                 \frac{u_{\!f}}{A} & 0 \\
                 0 & \frac{u_b}{A} \\
               \end{array} \right) .
\end{eqnarray}
And the terms in $\overline{\sigma^y_L}$ given by the off-diagonal
elements is written out in the Appendix.

\begin{figure}[h]
\bigskip\bigskip
\includegraphics[width=76mm]{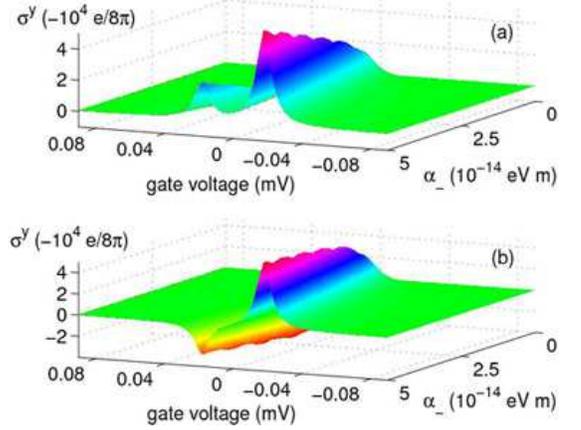}
\caption{(Color online) $\overline{\sigma^y}$ as function of $V$ and
$\alpha_-$. The relaxation-time differences are $\Delta\tau=8\times
10^{-5}$ in panel (a) and $\Delta\tau=5\times 10^{-4}$ in panel (b)
with $\tau_\textrm{b}=660~ \textrm{fs}$.}\label{fig:diff}
\end{figure}

The influence of the difference of impurity potentials on the tunneling spin
conductivity can be observed in Fig.~\ref{fig:diff}.
Here we introduce the difference in relaxation times
$\Delta\tau =(\tau_{\!f}-\tau_b)/\tau_b$
which is taken to be $8\times 10^{-5}$ and $5\times 10^{-4}$
in Fig.~\ref{fig:diff}(a) and Fig.~\ref{fig:diff}(b), respectively.
As the difference $\Delta\tau$ increases,
one of the resonant peak in $\overline{\sigma^y}$ tends
to be suppressed and  finally it becomes a valley.

Figure~\ref{fig:diff} shows that the variation of the strengthes
of impurity potentials between layers leads to
the asymmetric dependence of
the TSC on the gate voltage
when the in-plane driven electric field is fixed.
This unilateral conduction
feature makes the bilayer system with different strengthes
of impurity potentials a candidate for the realization of
spin diode.

\section{Summary}\label{sec:summary}

We have studied the coherent TSC in the bilayer
system with spin-orbit coupling.
We first revisited the definition of the tunneling charge current
with the help of continuity-like equations in the bilayer system
since the tunneling between layers causes the nonconservation of
the density in each layer.
In additional to the conventional contribution,
the tunneling current contains those related to
the overlap of wavefunctions for different layers.
This is intuitional for us to define the coherent TSC.
We showed that the contributions of the wavefunction
overlaps for states with spin
parallel and anti-parallel to reference axis
have opposite signs.
Unlike the conventional tunneling Hamiltonian approach,
the tunneling strength in our study is not necessarily week
since it is treated as a part of the unperturbed Hamiltonian.
In the light that only a gate voltage can not induce a nonvanishing result,
we studied TSC in response to an in-plane electric field.
The spin conductivity was calculated in terms of Kubo formula
by taking account of nonmagnetic impurities.
We firstly investigated twin-layer systems
and then considered the effect caused by the difference
between the strengths of impurity potentials in different
layers.
Meanwhile, we developed the techniques dealing with the impurity-averaged
Green's function in the bilayer system.
We showed that there must exist nonvanishing off-diagonal elements
in the averaged Green's function even in the chiral representation
if the strengthes of impurity potentials in two layers is different.
Sharp cusps in the tunneling spin conductivity appear
near the null voltage
and are suppressed by the impurities.
We found that if the strength of impurity potential in one layer
is different from that in the other layer, the TSC
exhibits the asymmetrical feature with respect to the gate voltage.
This reveals that the spin diode can also be realized
in the bilayer system with different strengthes
of impurity potentials in different layers.

\acknowledgements

The work was supported by NSFC Grant No. 10674117 and partially by
PCSIRT Grant No. IRT0754.

\appendix

\section{Expressions for some coefficients and the matrices}

The concrete expressions for the matrix elements of the averaged
retarded Green's function is given by
\begin{widetext}
\begin{eqnarray}
R_{11} &=& \frac{\omega-\varepsilon_2- w_+
 -\frac{eV-\alpha_-k}{\lambda_{12}}w_-}
 {(\omega-\varepsilon_1)(\omega-\varepsilon_2)
 -w_+(2\omega-\varepsilon_1-\varepsilon_2)
 +(\omega_++\omega_-)(\omega_+-\omega_-)-2\omega_-(eV-\alpha_-k)},
                     \nonumber \\
R_{22} &=& \frac{\omega-\varepsilon_1- w_+
 +\frac{eV-\alpha_-k}{\lambda_{12}}w_-}
 {(\omega-\varepsilon_1)(\omega-\varepsilon_2)
 -w_+(2\omega-\varepsilon_1-\varepsilon_2)
 +(\omega_++\omega_-)(\omega_+-\omega_-)-2\omega_-(eV-\alpha_-k)},
                     \nonumber \\
R_{33} &=& \frac{\omega-\varepsilon_4- w_+
 -\frac{eV+\alpha_-k}{\lambda_{34}}w_-}
 {(\omega-\varepsilon_3)(\omega-\varepsilon_4)
 -w_+(2\omega-\varepsilon_3-\varepsilon_4)
 +(\omega_++\omega_-)(\omega_+-\omega_-)-2\omega_-(eV+\alpha_-k)},
                     \nonumber \\
R_{44} &=& \frac{\omega-\varepsilon_3- w_+
 +\frac{eV+\alpha_-k}{\lambda_{34}}w_-}
 {(\omega-\varepsilon_3)(\omega-\varepsilon_4)
 -w_+(2\omega-\varepsilon_3-\varepsilon_4)
 +(\omega_++\omega_-)(\omega_+-\omega_-)-2\omega_-(eV+\alpha_-k)},
                     \nonumber \\
R_{\textrm{f}} &=&
 -\frac{\frac{\beta}{\lambda_{12}}w_-}
 {(\omega-\varepsilon_1)(\omega-\varepsilon_2)
 -w_+(2\omega-\varepsilon_1-\varepsilon_2)
 +(\omega_++\omega_-)(\omega_+-\omega_-)-2\omega_-(eV-\alpha_-k)},
                     \nonumber \\
R_{\textrm{b}} &=&
 -\frac{\frac{\beta}{\lambda_{34}}w_-}
 {(\omega-\varepsilon_3)(\omega-\varepsilon_4)
 -w_+(2\omega-\varepsilon_3-\varepsilon_4)
 +(\omega_++\omega_-)(\omega_+-\omega_-)-2\omega_-(eV+\alpha_-k)},
\end{eqnarray}
with $w_\pm=\half((u_{\!f}\pm u_{b})n_\text{im}-\frac{i}{\tau_\pm})$
where $\tau_\pm=\half(\tau_{\!f}-\tau_b)$ are introduced.

When the strengthes of impurities are equal in the two layers, the
vertex correction to $\overline{\sigma^y}$ in terms of Green's
functions is given by
\begin{eqnarray}\label{eq:conduct-cross}
\overline{\sigma^y_L}=-
 \sum_{\mathbf k}\cos^2\!\varphi \Bigl\{\textrm{Re}
 \Bigl[(\frac{k}{m}\!-\!\alpha_+
 \!+\!\frac{\alpha_-(eV\!-\!\alpha_-k)}
 {\lambda_{12}}) a_1g_1
 \!+\!(\frac{k}{m}\!+\!\alpha_+
 \!-\!\frac{\alpha_-(eV\!-\!\alpha_-k)}
 {\lambda_{12}}) a_2g_2 \!-\!\alpha_+(a_1\!+\!a_2)(g_3\!+\!g_4)
        \hspace{3mm}    \nonumber \\
 -\!(\frac{k}{m}\!+\!\alpha_+
 \!-\!\frac{\alpha_-(eV\!+\!\alpha_-k)}
 {\lambda_{34}}) a_3g_3
 \!-\!(\frac{k}{m}\!+\!\alpha_+
 \!+\!\frac{\alpha_-(eV\!+\!\alpha_-k)}
 {\lambda_{34}}) a_4g_4
 \!-\!\alpha_+\frac{e^2V^2\!+\!\beta^2\!-\!\alpha^2_-k^2}
 {\lambda_{12}\lambda_{34}}(a_1\!-\!a_2)(g_3\!-\!g_4)
         \hspace{1.5mm}    \nonumber \\
 +\alpha_-(\frac{eV-\alpha_-k}{\lambda_{12}}
 (a_1-a_2)(g_3+g_4)
 +\frac{eV+\alpha_-k}{\lambda_{34}}
 (a_1+a_2)(g_3-g_4)) \Bigr]
 \frac{i(J_{12}+J_{34})}{2}
        \hspace{37mm}     \nonumber \\
 +\textrm{Re}\Bigl[
 -\frac{eV\!-\!\alpha_-k}{\lambda_{12}}
 ((\frac{k}{m}\!-\!\alpha_+
 \!+\!\frac{\alpha_-(eV\!-\!\alpha_-k)}{\lambda_{12}}) a_1g_1
 \!-\!(\frac{k}{m}\!-\!\alpha_+
 \!-\!\frac{\alpha_-(eV\!-\!\alpha_-k)}{\lambda_{12}}) a_2g_2
 -\alpha_+(a_1\!-\!a_2)(g_3\!+\!g_4))
              \nonumber \\
 \!+\!\frac{eV\!+\!\alpha_-k}{\lambda_{34}}
 ((\frac{k}{m}\!+\!\alpha_+
 \!-\!\frac{\alpha_-(eV\!+\!\alpha_-k)}{\lambda_{34}}) a_3g_3
 \!-\!(\frac{k}{m}\!+\!\alpha_+
 \!+\!\frac{\alpha_-(eV\!+\!\alpha_-k)}{\lambda_{34}}) a_4g_4
 \!+\!\alpha_+(a_1\!+\!a_2)(g_3\!-\!g_4))
        \hspace{10.5mm}           \nonumber \\
 -\alpha_-(a_1\!+\!a_2)(g_3\!+\!g_4)
 \!-\!2\beta^2\alpha_-
 (\frac{a_1g_2}{\lambda^2_{12}}\!+\!\frac{a_3g_4}{\lambda^2_{34}})
 \!-\!\frac{\alpha_-(e^2V^2\!-\!\beta^2\!-\!\alpha^2_-k^2)}
 {\lambda_{12}\lambda_{34}}(a_1\!-\!a_2)(g_3\!-\!g_4)
 \Bigr]\frac{i(J_{12}-J_{34})}{2}
        \hspace{10mm}       \nonumber \\
 +\textrm{Re}\Bigl[
 -(\frac{k}{m}-\alpha_+ +\frac{\alpha_-(eV-\alpha_-k)}
 {\lambda_{12}}) \frac{a_1g_1}{\lambda_{12}}
 +(\frac{k}{m}-\alpha_+
 -\frac{\alpha_-(eV-\alpha_-k)}
 {\lambda_{12}}) \frac{a_2g_2}{\lambda_{12}}
 +\frac{\alpha_+}{\lambda_{12}}(a_1-a_2)(g_3+g_4)
          \hspace{4.5mm}     \nonumber \\
 +(\frac{k}{m}+\alpha_+
 -\frac{\alpha_-(eV+\alpha_-k)}
 {\lambda_{34}}) \frac{a_3g_3}{\lambda_{34}}
 -(\frac{k}{m}+\alpha_+
 +\frac{\alpha_-(eV+\alpha_-k)}
 {\lambda_{34}}) \frac{a_4g_4}{\lambda_{34}}
 +\frac{\alpha_+}{\lambda_{34}}(a_1+a_2)(g_3-g_4)
          \hspace{13mm}     \nonumber \\
 +2\alpha_-(\frac{eV-\alpha_-k}{\lambda^2_{12}}a_1g_2
 +\frac{eV+\alpha_-k}{\lambda^2_{34}}a_3g_4
 -\frac{eV}{\lambda_{12}\lambda_{34}}
 (a_1-a_2)(g_3-g_4))
 \Bigr] \frac{i\beta(J_{14}+J_{32})}{2}
         \hspace{32mm}       \nonumber \\
 +\textrm{Im}\Bigl[\frac{a_1g_2}{\lambda_{12}}
 \!-\!\frac{(a_3\!+\!a_4)(g_1\!-\!g_2)}{2\lambda_{12}}
 \!+\!\frac{a_3g_4}{\lambda_{34}}
 \!-\!\frac{(a_1+a_2)(g_3-g_4)}{2\lambda_{34}}
 \!-\!\frac{\alpha_+k}{\lambda_{12}\lambda_{34}}
 (a_1\!-\!a_2)(g_3\!-\!g_4)
 \Bigr]\alpha_-\beta(J_{14}-J_{32}) \Bigr\},
          \hspace{4mm}
\end{eqnarray}
where $J_{12},J_{14},J_{32},J_{34}$ are the matrix elements of the
vertex $\mathcal J^y$ which can be obtained by solving
Eq.~(\ref{eq:transfer}).

The difference of strengthes of impurity potentials in each layer brings the
off-diagonal elements to the Green's function in the chiral representation.
The corresponding contribution to the vertex correction of the
conductivity is given by
\begin{eqnarray}
\overline{\sigma^y_{L,d}} =-
 \sum_{\mathbf k} \cos^2\!\varphi\Bigl\{
 \textrm{Re}\Bigl[A_\textrm{f}R_\textrm{f}(\frac{k}{m}-\alpha_+)
 -A_\textrm{b}R_\textrm{b}(\frac{k}{m}+\alpha_+)
 +\beta\alpha_-(R_{11}+R_{22}+R_{33}+R_{44})(
 \frac{A_\textrm{f}}{\lambda_{12}}
 +\frac{A_\textrm{b}}{\lambda_{34}})
              \hspace{14mm}    \nonumber \\
 -\frac{2\alpha_+}{\lambda_{12}\lambda_{34}}(
 A_\textrm{f}R_\textrm{b}(e^2V^2+\beta^2-\alpha^2_-k^2)
 +(A_\textrm{f}(R_{33}-R_{44})
 -A_\textrm{b}(R_{11}-R_{22}))\beta\alpha_-k)
 \Bigr]i(J_{12}+J_{34})
              \hspace{23mm}    \nonumber \\
 +\textrm{Re}\Bigl[
  -\frac{\beta}{\lambda_{12}}A_\textrm{f}
 ((R_{11}+R_{22})(\frac{k}{m}-\alpha_+)
 -(R_{33}+R_{44})\alpha_+)
 +\frac{\beta}{\lambda_{34}}A_\textrm{b}
 ((R_{33}+R_{44})(\frac{k}{m}+\alpha_+)
 +(R_{11}+R_{22})\alpha_+)
                  \nonumber \\
 +\frac{2\alpha_-}{\lambda_{12}}
 (A_\textrm{f}R_\textrm{b}
 \frac{e^2V^2-\beta^2-\alpha^2_-k^2}{\lambda_{34}}
 -(A_\textrm{f}(R_{33}-R_{44})
 +A_\textrm{b}(R_{11}-R_{22}))\frac{eV\beta}{\lambda_{34}}
 +A_\textrm{f}R_\textrm{f}
 \frac{(eV-\alpha_-k)^2-\beta^2}{2\lambda_{12}})
              \hspace{9.5mm}    \nonumber \\
 -A_\textrm{f}(R_{11}\!-\!R_{22})
 \frac{2\alpha_-\beta(eV\!-\!\alpha_-k)}{\lambda^2_{12}}
 +A_\textrm{b}R_\textrm{b}
 \frac{\alpha_-((eV\!+\!\alpha_-k)^2\!-\!\beta^2)}{\lambda^2_{34}}
 \!-\!A_\textrm{b}(R_{33}\!-\!R_{44})
 \frac{2\beta(eV\!+\!\alpha_-k)}{\lambda^2_{34}}
 \Bigr]i(J_{12}-J_{34})
             \hspace{1mm}    \nonumber \\
 +\textrm{Re}\Bigl[
 \frac{eV\!-\!\alpha_-k}{\lambda_{12}}A_\textrm{f}
 ((R_{11}\!+\!R_{22})(\frac{k}{m}\!-\!\alpha_+)
 \!-\!\alpha_+(R_{33}\!+\!R_{44})
 \!+\!R_\textrm{f}\frac{2\alpha_-\beta}{\lambda_{12}})
 \!+\!\alpha_-A_\textrm{f}(R_{11}\!-\!R_{22})
 \frac{(eV\!-\!\alpha_-k)^2\!-\!\beta^2}{\lambda^2_{12}}
             \hspace{6mm}    \nonumber \\
 +\frac{eV\!+\!\alpha_-k}{\lambda_{34}}A_\textrm{b}
 ((R_{33}\!+\!R_{44})(\frac{k}{m}\!+\!\alpha_+)
 \!-\!\alpha_+(R_{11}\!+\!R_{22})
 \!+\!R_\textrm{b} \frac{2\alpha_-\beta}{\lambda_{34}})
 \!+\!\alpha_-A_\textrm{b}(R_{33}\!-\!R_{44})
 \frac{(eV\!+\!\alpha_-k)^2\!-\!\beta^2}{\lambda^2_{34}}
             \hspace{9.5mm}     \nonumber \\
 +\frac{\alpha_-}{\lambda_{12}\lambda_{34}}
 (4eV\beta A_\textrm{f}R_\textrm{b}
 +(A_\textrm{f}(R_{33}-R_{44})+A_\textrm{b}(R_{11}-R_{22}))
 (e^2V^2-\beta^2-\alpha^2_-k^2))
 \Bigr]i(J_{14}+J_{32})
             \hspace{23.5mm}     \nonumber \\
 -\textrm{Im}\Bigl[
 \frac{\alpha_+}{\lambda_{12}\lambda_{34}}
 (4\beta\alpha_-k A_\textrm{f}R_\textrm{b}
 \!-\!(A_\textrm{f}(R_{33}\!-\!R_{44})
 \!-\!(A_{11}\!-\!A_{22})R_\textrm{b})
 (e^2V^2\!+\!\beta^2\!-\!\alpha^2_-k^2))
 \!+\!A_\textrm{f}(R_{11}\!-\!R_{22})(\frac{k}{m}\!-\!\alpha_+)
             \hspace{4mm}        \nonumber \\
 +\alpha_-(R_{11}\!+\!R_{22}\!+\!R_{33}\!+\!R_{44})(
 \frac{(eV\!-\!\alpha_-k)A_\textrm{f}}{\lambda_{12}}
 \!+\!\frac{(eV\!+\!\alpha_-k)A_\textrm{b}}{\lambda_{34}})
 \!-\!A_\textrm{b}(R_{33}\!-\!R_{44})(\frac{k}{m}\!+\!\alpha_+)
 \Bigr](J_{14}-J_{32})  \Bigr\}.
             \hspace{10.5mm}
\end{eqnarray}

\end{widetext}

\end{document}